\documentclass[showpacs,prb,reprint,preprintnumbers,amsmath,amssymb,superscriptaddress,aps]{revtex4-1}
\usepackage{graphicx}
\usepackage{dcolumn}
\usepackage{bm}

\begin{document}
\title{Stark effect, polarizability and electroabsorption in silicon nanocrystals}

\author{Ceyhun \surname{Bulutay}}
\email{bulutay@fen.bilkent.edu.tr}
\affiliation{Department of Physics, Bilkent University, Ankara 06800, Turkey}
\author{Mustafa \surname{Kulakci}}
\author{Ra\c{s}it \surname{Turan}}
\affiliation{Department of Physics, Middle East Technical University, Ankara 06531, Turkey.}

\begin{abstract}
Demonstrating the quantum-confined Stark effect (QCSE) in silicon nanocrystals (NCs) embedded 
in oxide has been rather elusive, unlike the other materials. Here, the recent experimental data 
from ion-implanted Si NCs is unambiguously explained within the context of QCSE using an 
atomistic pseudopotential theory. This further reveals that the majority of the Stark shift comes 
from the valence states which undergo a level crossing that leads to a nonmonotonic 
radiative recombination behavior with respect to the applied field. The polarizability of 
embedded Si NCs including the excitonic effects is extracted over a diameter  range 
of 2.5--6.5~nm, which displays a cubic scaling, 
$\alpha=c D_{\mbox{\begin{scriptsize}NC\end{scriptsize}}}^3$, with $c=2.436\times 10^{-11}$ C/(Vm), 
where $D_{\mbox{\begin{scriptsize}NC\end{scriptsize}}}$ is the NC diameter. 
Finally, based on intraband electroabsorption analysis, it is predicted that $p$-doped Si NCs 
will show substantial voltage tunability, whereas $n$-doped samples should be almost insensitive.
Given the fact that bulk silicon lacks the linear electro-optic effect as being a centrosymmetric 
crystal, this may offer a viable alternative for electrical modulation using $p$-doped Si NCs.
\end{abstract}
\pacs{73.22.-f, 78.67.Bf, 71.70.Ej}


\maketitle 

\section{Introduction}
The Stark effect has evolved within the previous century into a powerful spectroscopic tool for the
solids.\cite{macfarlane} In the case of bulk semiconductors, shallow free excitons become easily 
ionized which limits the strength of the applied fields. A more robust variant of the Stark effect
is the so-called quantum-confined Stark effect (QCSE) where the carriers are trapped in a quantum 
well or a lower dimensional structure.\cite{miller84} In this regard, the quantum dots or 
nanocrystals (NCs) are preferred so as to take advantage of the full three-dimensional 
confinement.\cite{bimberg} 
As a matter of fact, the electroabsorption studies were initiated quite early with CdS$_x$Se$_{1-x}$ 
NCs embedded in a glass matrix.\cite{rockstad} The QCSE activity in group-II-VI 
NCs\cite{hache89,colvin92,empedocles97} was soon extended to group-III-As NCs.\cite{heller,raymond} 
A related noteworthy  achievement was registering photoluminescence (PL) from a single quantum dot within an 
ensemble,\cite{nagamune} followed by probing the QCSE from a single dot.\cite{seufert01} 

On the technologically important front of group-IV materials, a recent breakthrough was the announcement 
of QCSE in germanium multiple quantum wells sandwiched between SiGe barrier layers.\cite{kuo05}
The drawback of this structure is the small band offset of the barrier regions which limits the applied 
reverse bias before carrier tunneling sets in. Furthermore, it suffers from the polarization-dependent response
discriminating between TE and TM polarizations; both of these shortcomings are inherently carried over 
to Si/Ge self-assembled quantum dots.\cite{elkurdi05}
Si NCs embedded in oxide, not only offer remedy to both of these problems, but also due to its insulating host matrix 
it can withstand very high electric fields. Surprisingly, even though nanosilicon has become an established 
field,\cite{daldosso} the QCSE activity in this system has been quite overlooked.
In some of the early electroluminescence and photoluminescence studies on Si NCs, 
the precursors of QCSE was reported as a small redshift which was however taken 
over by a strong blueshift.\cite{photopoulos,ioannou}
As another indirect measurement of QCSE in Si NCs, Lin \emph{et al.} announced a 11~nm redshift 
under strong illumination, but without an external bias, which they attributed to a 
build up of an internal electric field due to capture of 
carriers in NCs.\cite{lin04} Only very recently, the direct measurement of QCSE under an 
external field in Si NCs was achieved 
on ion-implanted samples  that yielded as large as a 40~nm redshift at cryogenic 
temperatures, which remained to be easily detectable at the room temperature.\cite{kulakci08}  
This much delayed progress may nevertheless become crucial for the electronically controllable 
silicon-based photonics and especially for optical modulators; the latter has been a real challenge, 
as bulk silicon, being a centrosymmetric crystal, lacks the Pockels effect which leaves the 
plasma effect as the main route for electrical 
modulation.\cite{reed06,soref06} Very recently, a GeSi electroabsorption modulator has been 
announced that makes use of the bulk Franz-Keldysh effect of germanium enhanced under tensile 
strain.\cite{liu08} Amidst these developments, the present understanding on the QCSE and 
electroabsorption in Si NCs remains to be quite insufficient so as to address whether it can offer a viable 
alternative to the existing and emerging ones.

In this work, we aim for an assessment of these prospects from a rigorous atomistic point of view,
starting with the recent QCSE experiment and extending our analysis to both 
fundamental as well as applied directions. First, we theoretically show that, the highly pronounced 
luminescence shifts as measured in Ref.~\onlinecite{kulakci08} unambiguously originates 
from the Stark effect. In so doing, the importance of the excitonic effects is emphasized for 
larger NCs. The detailed explanation of the emission strength as a function of Stark field reveals 
the intricate interplay of the single-particle Stark shifts, their level crossings and the electric field 
dependence of the oscillator strengths. From a fundamental point of view, in this context the 
most important physical quantity is their polarizability. 
However, this is a subject which has not been discussed so far in the literature. 
Therefore, we provide the polarizability of 
embedded Si NCs for the useful 2.5--6.5~nm diameter range and furthermore, our exhaustive 
computations are expressed in simple expressions to enable their use by other researchers.
Finally, we complete our theoretical analysis with the intraband electroabsorption properties 
of Si NCs under a strong electric field, where we predict the marked discrepancy 
between the $n$- and $p$-doped NCs.

Our computational framework is a semiempirical pseudopotential-based atomistic Hamiltonian\cite{bester09} 
in conjunction with the linear combination of bulk bands (LCBB) as the expansion basis.\cite{ninno,zunger} 
The strong Stark field is included directly to the Hamiltonian without any perturbative approximation.
This is the current state-of-the-art theory for this system size, which is far advanced compared to 
effective mass and envelope functions approaches\cite{efros96,ouisse00,desousa05} 
(for a critical account of the latter, see Ref.~\onlinecite{wood}) 
and moreover is not amenable by {\em ab initio} techniques.\cite{note1}
The competence of this technique has been well tested; in the case of embedded Si and Ge NCs, 
this has been employed to study interband and intraband optical
absorptions,\cite{bulutay07} the Auger recombination, and carrier multiplication;\cite{sevik08} 
furthermore, predictions for the third-order nonlinear optical properties using the same 
theoretical approach\cite{yildirim08} have been independently verified experimentally.\cite{imakita09}
The paper is organized as follows. A description of the theoretical method is given in Sec.~II.
The QCSE, polarizability, and intraband electroabsorption are analyzed in Sec.~III. 
Main conclusions are presented in Sec.~IV.

\section{Theory}
In the LCBB approach,\cite{zunger} the NC wave function with a state index $j$ is expanded
in terms of the bulk Bloch band ($n$) and the wave vector ($\vec{k}$) as,
\begin{equation}
\psi_j(\vec{r})=\frac{1}{\sqrt{N}}\sum_{n,\vec{k},\mu} C^{\mu}_{n,\vec{k},j}\,
e^{i\vec{k}\cdot\vec{r}} u^{\mu}_{n,\vec{k}}(\vec{r})  \, ,
\end{equation}
where $N$ is the number of primitive cells within the computational supercell,
$C^{\mu}_{n,\vec{k},j}$ is the expansion coefficient set to be determined,
and $\mu$ is the constituent bulk material label pointing to the NC core and
embedding medium. $u^{\mu}_{n,\vec{k}}(\vec{r})$ is the cell-periodic part
of the Bloch states which can be expanded in terms of the reciprocal-lattice
vectors, $\{\vec{G}\}$ as
\begin{equation}
 u^{\mu}_{n,\vec{k}}(\vec{r})=\frac{1}{\sqrt{\Omega_0}}\sum_{\vec{G}}
 B^{\mu}_{n\vec{k}}\left(\vec{G}\right)e^{i\vec{G}\cdot\vec{r}}\, ,
\end{equation}
where $\Omega_0$ is the volume of the primitive cell.
The Hamiltonian has the usual kinetic and the ionic potential parts, the 
latter for describing the atomistic environment within the pseudopotential framework,
given by
\begin{eqnarray}
\label{Hamiltonian}
\hat{H} & = & \hat{T}+\hat{V}_{\mbox{\begin{scriptsize}PP\end{scriptsize}}} \nonumber\\ & = & 
-\frac{\hbar^2\nabla^2}{2m_0}+
\sum_{\mu,\vec{R}_j,\alpha} W^{\mu}_{\alpha}(\vec{R}_j)\,
\upsilon^{\mu}_{\alpha}\left( \vec{r}-\vec{R}_j-\vec{d}^{\mu}_{\alpha}\right) \, ,
\end{eqnarray}
where $m_0$ is the free electron mass,
$W^{\mu}_{\alpha}(\vec{R}_j)$ is the atomic identity coefficient that takes values
0 or 1 depending on the type of atom at the position 
$\vec{R}_j-\vec{d}^{\mu}_{\alpha}$, here $\vec{R}_j$ is the primitive cell coordinate and 
$\vec{d}^{\mu}_{\alpha}$ is the displacement of this particular atom within the primitive cell.
$\upsilon^{\mu}_{\alpha}$ is the screened spherical pseudopotential
of atom $\alpha$ of the material $\mu$, the latter distinguishes the NC and the 
matrix regions.

The formulation can be cast into the following generalized eigenvalue equation,\cite{zunger,chirico}
\begin{equation}
\sum_{n,\vec{k},\mu}H_{n'\vec{k}'\mu',n\vec{k}\mu}\,
C^{\mu}_{n,\vec{k},j} =E_j \sum_{n,\vec{k},\mu}S_{n'\vec{k}'\mu',n\vec{k}\mu}\,C^{\mu}_{n,\vec{k},j} \, ,
\end{equation}
where,
$$
H_{n'\vec{k}'\mu',n\vec{k}\mu}  \equiv  \left\langle n'\vec{k}
'\mu'\vert\hat{T}+\hat{V}_{\mbox{\begin{scriptsize}PP\end{scriptsize}}}\vert
n\vec{k}\mu \right\rangle\, ,
$$
$$
\left\langle n'\vec{k}
'\mu'\vert\hat{T}\vert n\vec{k}\mu \right\rangle  =  \delta_{\vec{k}',\vec{k}}
\sum_{\vec{G}}\frac{\hbar^2}{2m}\left\vert \vec{G}+\vec{k} \right\vert^2
B^{\mu'}_{n'\vec{k}'}
\left(\vec{G}\right)^*
B^{\mu}_{n\vec{k}}\left(\vec{G}\right)\, ,
$$
\begin{eqnarray}
\left\langle n'\vec{k}
'\mu'\vert\hat{V}_{\mbox{\begin{scriptsize}PP\end{scriptsize}}}\vert n\vec{k}\mu
\right\rangle & = & \sum_{\vec{G},\vec{G}'}
 B^{\mu'}_{n'\vec{k}'} \left(\vec{G}'\right)^* B^{\mu}_{n\vec{k}}\left(\vec{G}\right)
\nonumber \\ & & \times
 \sum_{{\mu''},\alpha}V_\alpha^{{\mu''}}\left( \left\vert \vec{G}+\vec{k}-\vec{G}
 '-\vec{k}'\right\vert^2\right)
\nonumber \\ & & \times
 W_\alpha^{{\mu''}}\left(\vec{k}-\vec{k}'\right)
 e^{i\left(\vec{G}+\vec{k}-\vec{G}'-\vec{k}'\right)\cdot\vec{d}_\alpha^{{\mu''}}}
 \, , \nonumber
\end{eqnarray}
the overlap part in the generalized eigenvalue equation is of the form
$$
S_{n'\vec{k}'\mu',n\vec{k}\mu}  \equiv
\left\langle n'\vec{k}
'\mu'\vert n\vec{k}\mu \right\rangle\, .
$$
The Si NC is intended to be embedded in silica, represented by an artificial 
wide band gap host matrix that has the same band-edge line up and the dielectric constant, 
but otherwise lattice-matched with the diamond structure of Si.\cite{bulutay07}
We refer to Ref.~\onlinecite{bulutay07} for the other technical details of the implementation 
of the electronic structure, including the form of the pseudopotentials for the NC and matrix media.

\begin{figure}
\includegraphics[width=9 cm, angle=0]{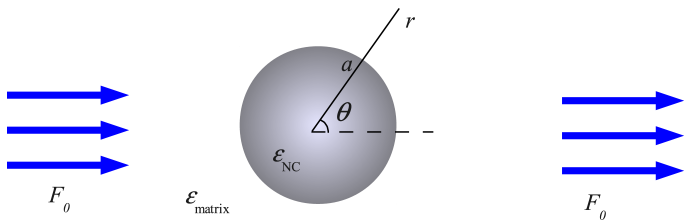}
\caption{\label{fig-qcse-cartoon}
(Color online) The electrostatic setting of the embedded NC under a dc external field.}
\end{figure}

For the study of QCSE, we treat the strong external field in the same level as the other
terms of the atomistic Hamiltonian (i.e., nonperturbatively).
At variance with the electrostatic model used in Ref.~\onlinecite{kulakci08}, we assume that an
individual spherical Si NC under consideration is embedded in a \emph{uniform} medium having a  
constant permittivity for silicon rich oxide. This is justified by the spatial distribution of the 
light-emitting centers in Si-implanted SiO$_2$.\cite{serincan}
In connection to the actual samples, we inherently assume that the NCs are well separated 
which applies safely to NC volume filling factors of about 10\% or less. 
The basic electrostatic construction of the
problem is presented with the assumption that the NCs are well separated in Fig.~\ref{fig-qcse-cartoon}.
If we denote the uniform applied electric field in the matrix region asymptotically away from the NC as
$F_{\mbox{\begin{scriptsize}0\end{scriptsize}}}$, then the
solution for electrostatic potential is given in spherical coordinates by~\cite{jackson}
\begin{equation}
\label{phi-ext}
\Phi(r,\theta)=\left\{
\begin{array}{ll}
         -\frac{3}{\epsilon+2}F_{\mbox{\begin{scriptsize}0\end{scriptsize}}} r \cos\theta\, , & r \leq a \\
        -F_{\mbox{\begin{scriptsize}0\end{scriptsize}}} r \cos\theta+\left( \frac{\epsilon -1}{\epsilon+2}\right)
          F_{\mbox{\begin{scriptsize}0\end{scriptsize}}}\frac{a^3}{r^2}\cos\theta\, ,
         & r > a\end{array}
 \right. ,
\end{equation}
where $\epsilon \equiv\epsilon_{\mbox{\begin{scriptsize}NC\end{scriptsize}}}
/\epsilon_{\mbox{\begin{scriptsize}matrix\end{scriptsize}}}$
is the ratio of the permittivities of the inside and outside of the NCs.
Hence, this expression accounts for the surface polarization effects due to dielectric 
inhomogeneity which partially screens the external (i.e., dc Stark) field.
The effect of this external field can be incorporated by adding the
$V_{\mbox{\begin{scriptsize}ext\end{scriptsize}}}=e\Phi$ term to the potential-energy matrix elements.
A computationally convenient recipe in the context of QCSE is to assume that external 
potential is relatively smooth so that its
Fourier transform can be taken to be band limited to the first Brillouin zone of the underlying unit
cell.\cite{chirico} 
This results in the following LCBB matrix element
\begin{eqnarray}
\left\langle n'\vec{k}
'\mu'\vert\hat{V}_{\mbox{\begin{scriptsize}ext\end{scriptsize}}}\vert n\vec{k}\mu
\right\rangle & = & e\Phi(\vec{k}-\vec{k}') \sum_{\vec{G},\vec{G}'}
B^{\mu'}_{n'\vec{k}'} \left(\vec{G}'\right)^*  B^{\mu}_{n\vec{k}}\left(\vec{G}\right)
\nonumber \\ & & \times
\mbox{Rect}_{\vec{b}_1,\vec{b}_2,\vec{b}_3}(\vec{G}+\vec{k}-\vec{G}'-\vec{k}'),
\end{eqnarray}
here $\Phi(\vec{k})$ is the Fourier transform of $\Phi(\vec{r})$ as given in Eq.~(\ref{phi-ext}), 
$\mbox{Rect}_{\vec{b}_1,\vec{b}_2,\vec{b}_3}$ is the rectangular pulse function, which yields unity when
its argument is within the first Brillouin zone defined by the reciprocal-lattice
vectors, \{$\vec{b}_1,\vec{b}_2,\vec{b}_3$\}, and zero otherwise.

As will be supported by our following results, concomitant with the Stark redshift of the 
single-particle energies, the segregation of the electron and hole wave functions gives rise to a 
blueshift that partially negates the QCSE. To account for this effect, we include perturbatively\cite{note2} 
the so-called diagonal direct Coulomb term, $J_{vv,cc}$ between the valence-state wave 
function, $\psi_v(\vec{r})$, and the conduction-state wave function, $\psi_c(\vec{r})$, 
using the expression
\begin{equation}
J_{vv,cc}=-\int d^3r_1 d^3r_2 \frac{e^2 \left| \psi_c(\vec{r_1}) \right|^2 
\left| \psi_v(\vec{r_2}) \right|^2}{\epsilon(\vec{r_1},\vec{r_2})\left| \vec{r_1}-\vec{r_2}\right|},
\end{equation}
where, $e$ is the electronic charge, $1/\epsilon(\vec{r_1},\vec{r_2})$ is 
the inhomogeneous inverse dielectric function for which we use 
the mask function approach of Ref.~\onlinecite{wang03}. 
This is the only, but by far the dominant Coulomb term that is included in this study. 
A more elaborate approach for the electrostatic as well as the nondiagonal Coulomb 
terms can be found in Ref.~\onlinecite{franceschetti00}.
Furthermore, we ignore the spin-orbit coupling and hence, the NC states in this work are doubly 
spin degenerate. This coupling is particularly weak in silicon with its small atomic number and 
as a matter of fact, it forms the basis for spin-based silicon quantum computing proposals.\cite{fodor06}
Accordingly, no spin-flip process is considered in the carrier relaxation following 
the optical excitation so that only spin-triplet excitons are formed for which 
there is no exchange Coulomb contribution. 
Even if spin flips were to be allowed, for the NC size ranges of this study, their contribution 
which decay with the third power of diameter\cite{efros96} would become totally negligible compared to
the direct Coulomb and the single-particle Stark energies. Obviously, future studies can 
avoid some of these simplifications in this work.
  
\section{Results}
\subsection{Stark effect}
\begin{figure}
\includegraphics[width=9cm,clip]{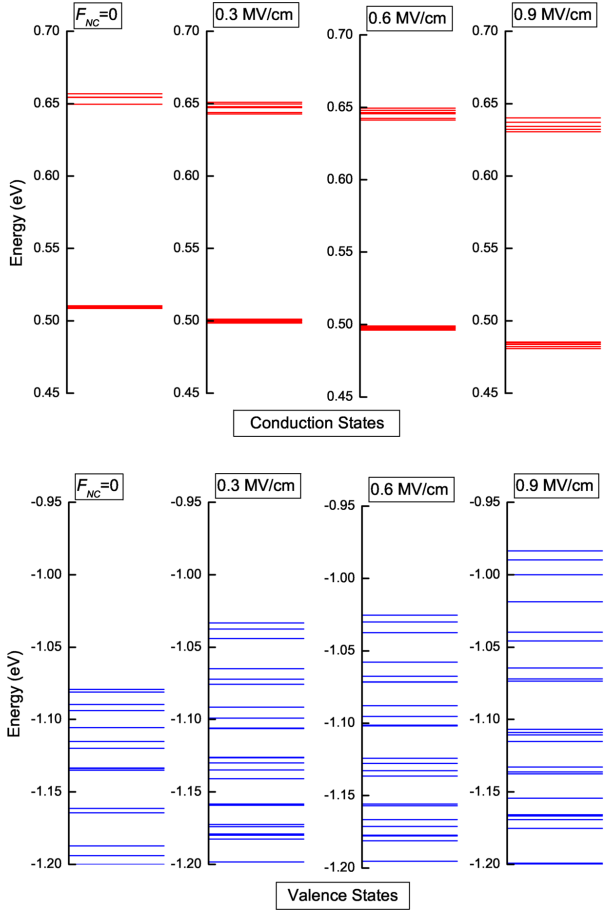}
\caption{\label{bar-spectra}(Color online) Single-particle energy levels of a 5.6~nm diameter Si NC 
for different internal NC electric fields.}
\end{figure}

\begin{figure}
\includegraphics[width=9cm,clip]{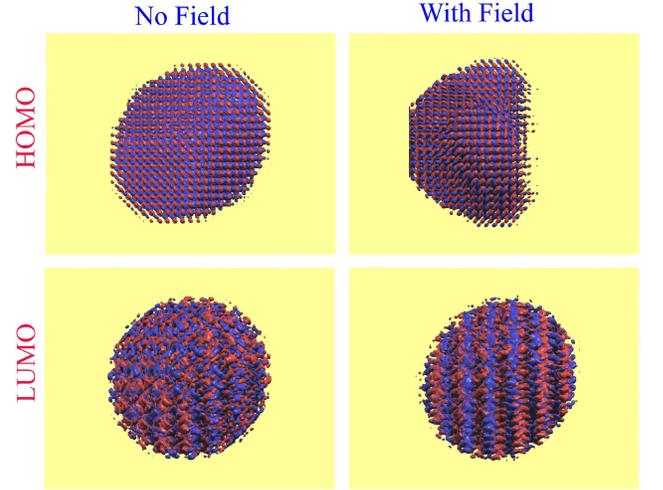}
\caption{\label{wf-iso}(Color online) HOMO (upper row) and LUMO (lower row) wave function isosurface
profiles of a 5.6~nm diameter Si NC under no (left column)  and 0.6~MV/cm (right column)
internal electric fields. The opposite signs of the wave function are 
represented by blue (dark) and red (light) colors. The electric field is horizontally directed 
from right to left. }
\end{figure}

In the recent experimental demonstration of QCSE in Si NCs,\cite{kulakci08} the wavelength for the 
peak-emitted intensity occurs at 780~nm. Based on our prior theoretical study,\cite{bulutay07} 
the corresponding diameter of the NC that matches with this optical gap is extracted as 5.6~nm.
For this size of a NC, we display in Fig.~\ref{bar-spectra} the evolution of the single-particle states with applied
Stark field. This clearly reveals that valence states are more prone to Stark shifts which was also 
observed to be the case in InP quantum dots.\cite{fu02} Indeed, Fig.~\ref{wf-iso} vividly demonstrates
that the highest-occupied molecular orbital (HOMO) 
wave-function distribution is significantly shifted by the Stark field and gets spatially 
squeezed between the high Stark field and the spherical NC interface. 
On the other hand, the lowest-unoccupied molecular orbital (LUMO) state encounters only a slight displacement, 
which is in the opposite direction with respect to HOMO as expected. 
According to stronger confinement of the valence states under 
the electric field, the interlevel separations become wider than those in the conduction states, 
as can be checked from Fig.~\ref{bar-spectra}. However, it needs to be reminded that the 
size quantization energy of the electrons is much larger than holes in Si NCs.\cite{note3}

\begin{figure}
\includegraphics[width=9 cm, angle=0]{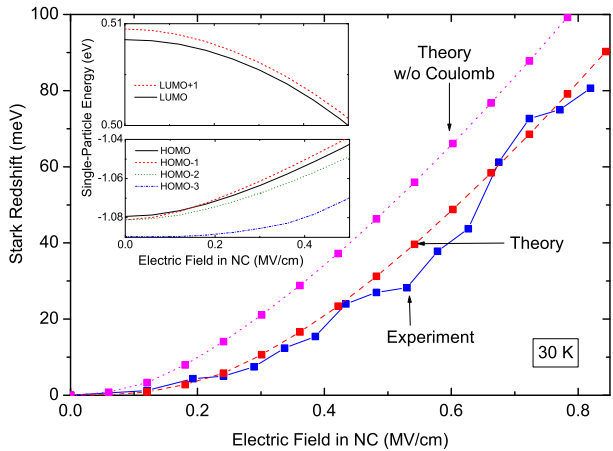}
\caption{\label{comp-stark}
(Color online) The comparison of theoretical and experimental\cite{kulakci08} Stark redshifts of a 
5.6~nm diameter Si NC at 30~K. 
The dotted line shows the theoretical curve without the direct Coulomb term included. 
Lines are solely for guiding the eye. Inset shows the single-particle Stark shifts of the band-edge 
states for the conduction (upper panel) and valence  states (lower panel).}
\end{figure}

In Fig.~\ref{comp-stark} we compare the 
experimental Stark redshift data\cite{kulakci08} at 30~K with our present theoretical results.   
To correlate with this PL experiment and account for the thermal effects 
as well as the brightness of each excited-state recombination, 
we use the following Boltzmann factor-averaged and oscillator strength-averaged radiative recombination 
(i.e., emission) energy 
\begin{equation}
\bar{E}_{\mbox{\begin{scriptsize}emission\end{scriptsize}}}=\frac{\sum_{c,v} 
E_{cv} e^{-\beta (E_{cv}-E_{LH})}f_{cv}}
{\sum_{c,v} e^{-\beta (E_{cv}-E_{LH})}f_{cv}},
\end{equation}
here, $E_{cv}=E_c-E_v$, $E_{LH}$ are the conduction ($c$) to valence ($v$) state and 
LUMO to HOMO transition energies, respectively; $f_{cv}$ is the Cartesian-averaged oscillator 
strength of the transition\cite{bulutay07} and $\beta=1/(kT)$ with $k$ being the 
Boltzmann constant. We apply exactly the same averaging on the direct diagonal 
Coulomb energy by replacing the first $E_{cv}$ term in the numerator with $J_{vv,cc}$. 
The necessity for the Coulomb term is justified by the excellent agreement 
with the experimental data in Fig.~\ref{comp-stark}, whereas without it (i.e., at the
single-particle level) Stark shifts become significantly overestimated.
Since our model does not incorporate any size inhomogeneity, interface states or 
the strain effects, its success also supports the atomistic quantum confinement 
framework as the main source of the luminescence in these particular Si NC samples.
Furthermore, we note that as in the experimental work,\cite{kulakci08} we do not observe 
a dipolar term that gives rise to a linear dependence to the electric field.

The inset of Fig.~\ref{comp-stark} shows the evolution of the valence and conduction single-particle 
states of a 5.6~nm-diameter Si NC with respect to dc electric field. 
It indicates that there exists a level crossing between the HOMO and the HOMO-1 states around an 
internal NC field of 150~kV/cm. 

\begin{figure}
\includegraphics[width=9 cm, angle=0]{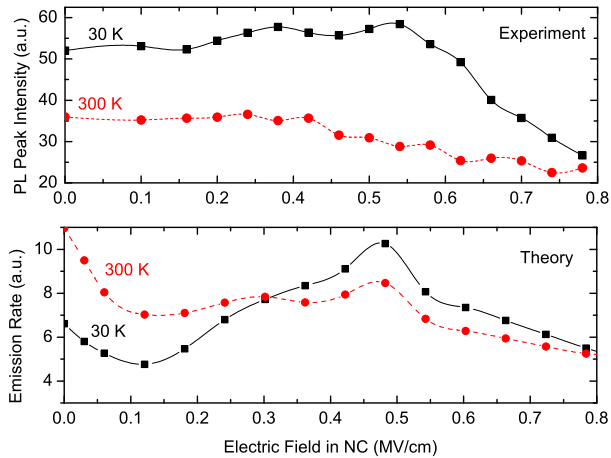}
\caption{\label{emission}
(Color online) The experimental PL peak intensity\cite{kulakci08} (upper plot) and the theoretical 
emission rate (lower plot) for Si NCs at 30 and 300~K.
Lines are solely for guiding the eye.}
\end{figure}

Figure~\ref{emission} contrasts the experimental PL peak intensity with the theoretical
radiative recombination rate, both as a function of applied electric field. The nonmonotonic 
behavior of the experimental data as well as the peak intensity appearing around 0.5~MV/cm
and the subsequent fall off beyond are all reproduced by the theory. 
However, for small Stark fields, the 300~K emission comes out to be stronger in the 
theoretical estimation. This may be due to thermally activated nonradiative processes 
that degrade the emission rate at higher temperatures. This deviation shows that further work is 
required to properly account for all thermal aspects of this problem.

\subsection{Polarizability of Si NCs}
\begin{figure}
\includegraphics[width=9 cm, angle=0]{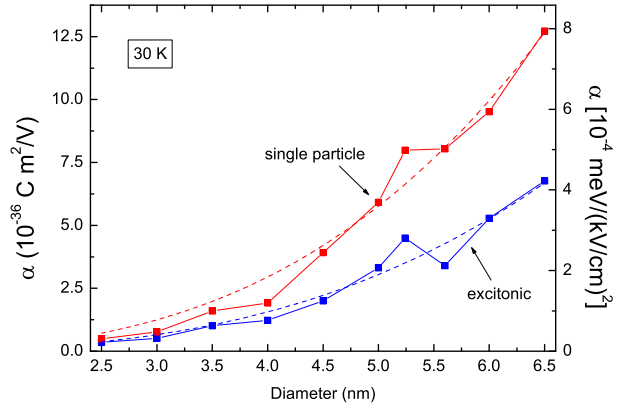}
\caption{\label{polarizability}
(Color online) Theoretically computed polarizability for Si NCs based on the single-particle and excitonic 
(i.e., with direct Coulomb term included) Stark shifts at 30~K. Solid lines are solely for guiding the eye.
The dashed lines are cubic fits to the data (see text).}
\end{figure}

The theoretical results up to now were restricted to a single diameter of 5.6~nm 
as extracted from the PL peak of the experimental data.\cite{kulakci08} We have ignored the 
size distribution of the NCs in the actual samples.\cite{serincan}
Next, we extract the size dependence of the polarizability of Si NCs, defined as $\Delta E=-(1/2)
\alpha F_{\mbox{\begin{scriptsize}NC\end{scriptsize}}}^2$, where $\Delta E$ is the overall Stark 
shift in the energy and $F_{\mbox{\begin{scriptsize}NC\end{scriptsize}}}$ is the electric field 
\textit{inside} the NC. Here, the polarizability, $\alpha$, is taken as scalar 
which is in general a rank-2 tensor, however, we have observed that the variation in Stark 
shift with respect to relative orientation of the electric field and the crystallographic planes 
of the NC or the $c$-axis of the $C_{3v}$ point group of the NCs,\cite{bulutay07} give rise 
to less than 10~meV changes for the highest applied fields.
In Fig.~\ref{polarizability} we show the excitonic polarizability (i.e., with direct Coulomb 
term included) at 30~K. For comparison purposes, the single-particle polarizability is also
provided where this estimate becomes highly exaggerated for larger diameters. 
Both of these curves display a nonmonotonic behavior with respect to size which exists 
in other physical properties as well; such variations occur as the states, such as HOMO and LUMO, 
acquire different representations of the $C_{3v}$ point group for different NC diameters.\cite{bulutay07}
As in the basic dipole polarizability, their overall trend can be easily fitted by a cubic 
dependence $\alpha=c D_{\mbox{\begin{scriptsize}NC\end{scriptsize}}}^3$, 
where $D_{\mbox{\begin{scriptsize}NC\end{scriptsize}}}$ 
is the NC diameter, and with $c=2.436\times 10^{-11}$ and $4.611\times 10^{-11}$ C/(Vm), 
for the excitonic and single-particle cases, respectively; in another unit system they are 
expressed as $c=1.521\times 10^{15}$ and $2.878\times 10^{15}$ meV/[(kV)$^2$ cm], respectively.

\begin{figure}
\includegraphics[width=9cm,clip]{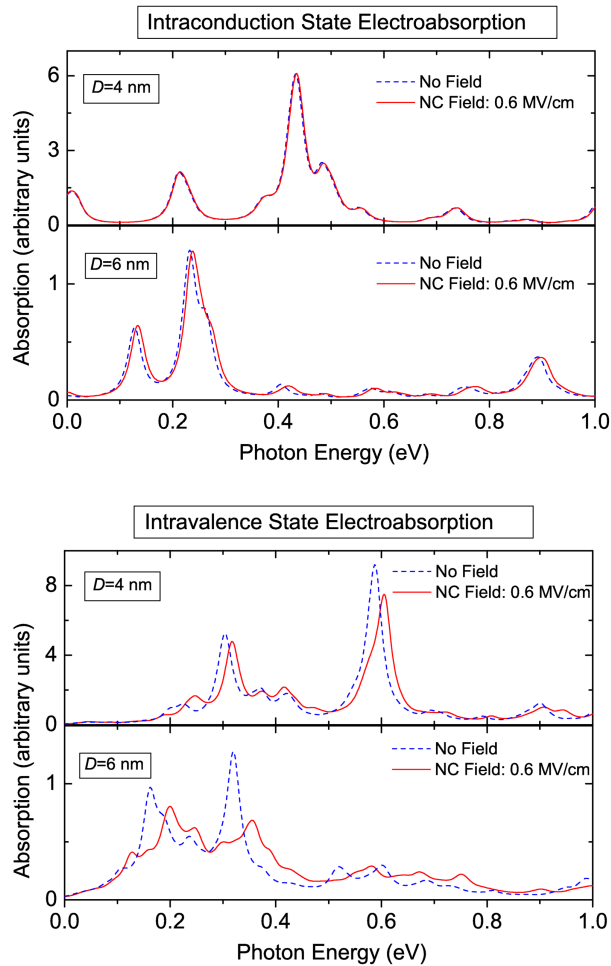}
\caption{\label{intersub-stark-abs}
(Color online) The intraconduction and intravalence state electroabsorption curves for 4~nm and 6~nm diameter Si NC at 300~K.
For amplitude comparison, all four curves are drawn to scale among themselves.}
\end{figure}

\subsection{Intraband electroabsorption of Si NCs}
In Fig.~\ref{intersub-stark-abs} the intravalence and intraconduction state electroabsorption
of 4~nm and 6~nm diameter Si NCs are shown, under 0 and 0.6~MV/cm internal NC electric fields. 
Using an anisotropic effective mass model and focusing only 
on the conduction band, de Sousa \emph{et al.}\cite{desousa05} have also modeled the intraband 
electroabsorption in Si NCs and as in this work, they obtained a blueshift.
However, as expected from the rigidity of the conduction states under the electric field 
(cf., Figs.~\ref{bar-spectra} and \ref{wf-iso}), and as observed in 
Fig.~\ref{intersub-stark-abs}, the electroabsorption effect can be best utilized in $p$-doped Si NCs. 
Unlike the interband transitions, we obtain a {\em blueshift} in the spectra with the applied electric field;
observe from Fig.~\ref{bar-spectra} that as the electric field increases, both the valence and 
conduction states individually fans out, whereas the optical band gap gets redshifted.
For the 6~nm case, the first peaks in the intravalence electroabsorption spectra 
shift close to 38~meV under a NC field of 0.6~MV/cm; this shift reduces to about 14~meV for the 4~nm case.
Given that bulk silicon has very poor Franz-Keldysh and Kerr effect efficiencies  for electro-optic 
modulation,\cite{soref87} these results can be encouraging for the consideration of nanocrystalline 
Si-based infrared electroabsorption modulators. However, we should also remark that the doping of 
Si NCs has its own challenges compared to bulk.\cite{ossicini05} 
Finally we should remark that as mentioned within the context of the Stark field,
the surface polarization effect, also known as local field effect, due to dielectric 
discontinuity, plays a role in the optical absorption as well.\cite{delerue03,ninno06,trani07} 
However, this effect becomes quite insignificant for Si NCs with diameters larger 
than about 4~nm and furthermore it only affects the amplitude of the absorption without 
modifying its spectral profile.\cite{bulutay07}

\section{Conclusion}
In conclusion, we show that the recent QCSE data\cite{kulakci08} for the embedded Si NCs under a 
strong Stark field can essentially be explained very well with an atomistic pseudopotential model 
which otherwise does not incorporate any size inhomogeneity, interface states, or the strain effects. 
In this context, the importance of the direct Coulomb interaction is demonstrated, and a simple expression 
for the Si NC polarizability is extracted. Finally, in compliance with the fact that the valence states 
display much more Stark shift, it is shown that intravalence band electroabsorption enjoys wider
voltage tunability which can be harnessed for Si-based electroabsorption modulator intentions.

\begin{acknowledgments}
The authors gratefully acknowledge the support by The Scientific and Technological Research Council of Turkey 
(T\"UB\.ITAK) under Projects No. 106T048 and No. 106M549. C.B. would like to thank Can U\u{g}ur Ayfer and the Bilkent 
University Computer Center for the critical computing service they provide.
\end{acknowledgments}


\begin{thebibliography}{99}
\bibitem{note3}We refer to Figs.~2 and 5 of Ref.~\onlinecite{bulutay07} which demonstrates the much 
faster shift of the LUMO with respect to NC size as compared to HOMO. Similar trend applies to all other conduction 
and valence states.
\end{thebibliography}
\end{document}